# Ni-O hybridization-driven electronic reconstruction across the superconducting dome in an infinite-layer nickelate

Chi Sin Tang[1,#,*], Shengwei Zeng[2,#], Xing Gao[3,#], Zhaoyang Luo[3], Xiongfang Liu[3,4], Zhi Shiuh Lim[2], Saurav Prakash[3], Ping Yang[1], Caozheng Diao[1], Xinmao Yin[4], Changjian Li[5], Huajun Liu[2], Mark B. H. Breese[1,3,*], A. Ariando[3,*]

[1]*Singapore Synchrotron Light Source (SSLS), National University of Singapore, 5 Research Link, Singapore 117603, Republic of Singapore.*

[2]*Institute of Materials Research and Engineering (IMRE), Agency for Science, Technology and Research (A*STAR), 2 Fusionopolis Way, Innovis #08-03, Singapore 138634, Republic of Singapore.*

[3]*Department of Physics, Faculty of Science, National University of Singapore, Singapore 117551, Republic of Singapore.*

[4]*Shanghai Key Laboratory of High Temperature Superconductors, Physics Department, Shanghai University, Shanghai 200444, China.*

[5]*Department of Materials Science and Engineering, Southern University of Science and Technology, Shenzhen 518055, Guangdong, China.*

[#]The authors contributed equally to this work.

*To whom correspondence should be addressed.

E-mail: slscst@nus.edu.sg; phymbhb@nus.edu.sg; ariando@nus.edu.sg

## Abstract

Superconductivity in infinite-layer nickelates has drawn wide interest as a cuprate analogue, yet how the electronic structure evolves with hole doping remains unsettled. Here we map the doping- and temperature-dependent unoccupied states of the La-based infinite-layer nickelate $La_{1-x}Ca_xNiO_2$ using O K-edge and Ni L-edge x-ray absorption spectroscopy. Superconductivity occurs for $0.18 \leq x \leq 0.27$. Near $x \approx 0.20 - 0.23$, low-energy spectral weight redistributes: Ni3$d$-dominated states decrease while O2$p$-hybridized states increase, indicating an orbital-selective crossover in Ni–O covalency. This crossover coincides with a sign reversal of the Hall coefficient and precedes the reduction of the superconducting critical temperature at higher doping. By directly linking transport anomalies and the superconducting dome to a measurable Ni–O orbital reorganization, our results provide a key step toward a unified, orbital-resolved phase diagram for infinite-layer nickelates and a practical route to engineer superconductivity via hybridization control.

## Introduction

The long-standing effort to understand unconventional superconductivity in cuprates has motivated the search for related superconductors in complex oxides. A major breakthrough was the discovery of superconductivity in Sr-doped $Nd_{1-x}Sr_xNiO_2$,[1] which positioned infinite-layer nickelates as promising platforms for high-$T_c$ research owing to their structural and electronic similarities to cuprates.[2-8] More recently, the superconducting critical temperatures in nickelates have come closer to those in the cuprates under high pressure[9-13] and also at ambient pressure.[14-17] Alongside these discoveries, advances in materials engineering, enabling not only stabilization of the infinite-layer phase[16] but also preservation of superconducting properties down to the atomic scale,[18] have accelerated progress in this field. These developments expand the experimental landscape for probing unconventional superconductivity under both extreme[19] and ambient conditions, and open new opportunities to study strongly correlated many-body systems across diverse tuning parameters.

Although superconducting nickelates share important features with cuprates, most notably quasi-two-dimensional electronic structures and strong local correlations,[1,2,5,20,21] they also differ in significant ways. In particular, both $3d_{x2-y2}$ and $3d_{z2}$ orbitals can contribute near the fermi surfaces,[4,22-26] in contrast to the predominantly single-band character of cuprates.[27] Infinite-layer nickelates also exhibit a distinct electronic landscape compared with canonical Mott insulators and other perovskite nickelates, including a Mott–Hubbard–like band structure and the absence of long-range antiferromagnetic order.[28,29] These differences complicate efforts to identify which electronic degrees of freedom are most relevant to superconductivity. Correspondingly, recent theoretical work has shifted attention away from rare-earth $4f$ orbitals and toward alternative interactions and orbital couplings.[30-33] A central open question is how the orbital character of low-energy states evolves with hole doping, specifically, whether Ni$3d$-dominated states give way to increasingly O$2p$-hybridized states near the compositions where superconductivity emerges and later weakens, and how such a crossover is reflected in transport signatures such as the sign change of the Hall coefficient.[2,21,34-36]

Here, using x-ray absorption spectroscopy (XAS) together with polarization-dependent x-ray linear dichroism (XLD), we track the evolution of O$2p$–Ni$3d$ hybridization in infinite-layer $La_{1-x}Ca_xNiO_2$ from the undoped to the overdoped regime. In contrast to earlier suggestions of weak O–Ni mixing, our measurements reveal a persistent and doping-dependent role of O$2p$–Ni$3d$ hybridization across the phase diagram. The elemental sensitivity of XAS provides direct access to unoccupied states above the Fermi level, while XLD offers orbital-resolved information, enabling a detailed view of how the electronic structure changes with doping and temperature.

From this systematic characterization, we identify superconductivity in the range $0.18 \leq x \leq 0.27$ in superconducting films of $La_{1-x}Ca_xNiO_2$. We find that near $x \approx 0.20$–$0.23$, low-energy spectral weight is transferred from Ni$3d$-dominated states to more strongly O-hybridized states, coincident with a polarity change in the Hall coefficient, $R_H$. This concurrence points to an orbital-selective reorganization of the low-energy electronic structure that goes beyond simple carrier counting. At higher doping, the continued reduction of Ni3$d$-dominated contributions and the increasing prominence of O-hybridized states track the suppression of $T_c$, linking the transport anomaly and the superconducting dome to a hybridization-driven evolution of the electronic structure. By clarifying how Ni–O orbital hybridization evolves with doping, this work refines the electronic phase diagram of $La_{1-x}Ca_xNiO_2$ and strengthens the case for nickelates as an alternative platform to cuprates for uncovering the ingredients of high-temperature superconductivity.

**Sample preparation and methods**

To systematically investigate how the electronic structure evolves with doping, we prepared and measured $La_{1-x}Ca_xNiO_2$ thin films spanning Ca concentrations $x$ from the undoped to the overdoped regime. Perovskite $La_{1-x}Ca_xNiO_3$ precursor films (thickness ~ 17 nm) were grown by pulsed laser deposition (PLD) without an $SrTiO_3$ (STO) capping layer, and subsequently converted to the infinite-layer phase via a soft-chemistry topotactic reduction process, following our previous report.[36] Details of the structural characterization and electrical transport measurements are provided there.

Temperature-dependent XAS measurements were carried out using linearly polarized x-rays at the soft x-ray–ultraviolet (SUV) beamline of the Singapore Synchrotron Light Source (SSLS). Spectra were collected in total electron yield (TEY) mode.

**Probing the electronic structures of $La_{1-x}Ca_xNiO_2$ via O K-edge XAS**

Unoccupied electronic states play a central role in determining a material's electronic structure and transport properties. These states can be probed by XAS at the oxygen and transition-metal absorption edges. Thus, we perform doping-dependent measurements across the infinite-layer series and compare the spectra with those of the corresponding perovskite precursor phase over the same Ca concentration range.

We first focus on the O K-edge ($1s \rightarrow 2p$), which is highly sensitive to hybridization between O2$p$ and transition-metal $3d$ states.[37] For nickelate thin films, the O K-edge therefore provides a direct probe of O2$p$–Ni3$d$ hybridization and, by extension, of the unoccupied Ni3$d$-derived states. In the perovskite phase $LaNiO_3$ and $La_{0.77}Ca_{0.23}NiO_3$, we observe a pronounced pre-peak at ~529.0 eV, commonly attributed to ligand-hole character on oxygen.[28,32,38] In contrast, this strong pre-peak disappears in the reduced infinite-layer films (Fig. 1(a), comparing as-grown perovskite $LaNiO_3$ with reduced infinite-layer $LaNiO_2$ at room temperature). The suppression of this feature is consistent with prior reports on infinite-layer $LaNiO_3$ and $NdNiO_2$ films,[28,32] and is typically associated with the filling of lower-lying $d$ states as the nominal Ni evolves from $3d^7$ toward $3d^9$ configurations.[32,39] This change is closely tied to the removal of apical oxygen and the accompanying reduction in ligand-hole density and Ni-O covalency.[22,40] A similar evolution is observed comparing perovskite $La_{0.77}Ca_{0.23}NiO_3$ with its reduced infinite-layer counterpart $La_{0.77}Ca_{0.23}NiO_2$ (Fig. S2(a)).

Notably, in the infinite-layer films we observe an additional, weaker pre-peak in this energy range (labelled "a" and indicated by the red arrow in Fig. 1(a)) at ~530.7 eV. This feature differs from many earlier reports where infinite-layer films were much thinner or capped with an STO layer.[2,41] Our films are relatively thick (~17 nm) and uncapped. Moreover, near-edge XAS measured in TEY mode has an effective probing depth of roughly ~5 nm,[42,43] supporting that feature "a" arises from electronic transitions intrinsic to the infinite-layer film. We therefore rule out a significant contribution from the STO substrate (see Supplementary Discussion). A similar pre-peak at this energy has been reported in infinite-layer $Nd_{0.8}Sr_{0.2}NiO_2$.[6] Related observations at $La_{2.85}Pr_{0.15}Ni_2O_7/SrLaAlO_4$ interfaces have also been interpreted as evidence for O2*p* states near $E_F$ and their hybridization with Ni3*d* orbitals, which may be relevant for superconductivity.[44]

Figure 1(b) summarizes the evolution of the O K-edge spectra across the full doping range $0 \leq x \leq 0.35$. In the pre-edge window (Region I, highlighted in yellow), feature "a" persists across the entire series, with a clear doping dependence. Starting from its minimum intensity in undoped $LaNiO_2$, the intensity of "a" increases with Ca substitution, saturating near the optimally doped composition ($La_{0.77}Ca_{0.23}NiO_2$), and then decreases slightly for higher doping levels ($x = 0.27$) and into the overdoped regime ($x = 0.35$). A similar trend is observed at low temperature (Fig. S1(b)).

In addition, as $x$ approaches 0.18 (near the onset of superconductivity), we observe the emergence of a second, weaker feature "b" at ~531.6 eV, just above feature "a". Feature "b" reaches its maximum intensity at $x = 0.23$ and gradually weakens for $x \geq 0.27$, coinciding with the overdoped regime where superconductivity diminishes. The same behaviour is observed at low temperature (Fig. S2(b)). We quantify these changes below via an analysis of the integrated spectral weight in Region I. Similar "satellite" or offshoot absorption features in the O K-edge pre-peak region have also been reported in optimally doped and overdoped cuprates.[45,46] Although their microscopic origin remains under debate, such features have been discussed in relation to changes in low-energy orbital character and stripe-related electronic reconstructions.[47,48]

At higher photon energies, beyond 534 eV (Region II, highlighted in green), we observe a shoulder "c" at ~534.8 eV, which can be assigned to hybridization between La5*d*/4*f* and O2*p* orbitals.[49-51] In the still higher-energy range (Region III, 536–547 eV; highlighted in blue), additional features are associated with hybridization between Ni4*sp*/La5*sp* and O2*p* states.[49-51] While these higher-energy features vary with doping, we do not observe a clear systematic trend.

Combining the XAS results with our prior transport measurements,[36] we propose a schematic, doping-dependent picture of the electronic structure of $La_{1-x}Ca_xNiO_2$ and contrast it with the perovskite precursor, as illustrated in Fig. 1(c–f). In undoped $LaNiO_2$, which is insulating below 300 K, feature "a" associated with O2*p*–Ni3*d* hybridized states lies just above the Fermi level $E_F$ but carries substantially less spectral weight than the corresponding unoccupied states in perovskite $LaNiO_3$ (Fig. 1(c,d)). With increasing Ca doping, the system becomes more metallic, and a superconductivity-associated spectral feature "b" emerges, as exemplified by $La_{0.77}Ca_{0.23}NiO_2$ (Fig. 1(e)). Upon further doping into the overdoped regime (e.g., $La_{0.65}Ca_{0.35}NiO_2$), feature "b" disappears and the metallicity of the infinite-layer phase weakens (Fig. 1(f)).

## Discussion

### Temperature-dependent O K-edge

Figure 2(a) shows representative O K-edge spectra of $La_{0.82}Ca_{0.18}NiO_2$ measured at 30 K and at room temperature (300 K). Overall, the main spectral features exhibit higher intensity at 300 K than at 30 K. This temperature dependence is consistently observed across the full doping range (see the low-temperature O K-edge spectra in Fig. S2(b)).

To quantify the temperature-induced changes in different photon-energy regions, we plot the temperature-dependent intensity difference, $\Delta I_T = I_{T=300\text{ K}} - I_{T=30\text{ K}}$, in Fig. 2(b). The results show that undoped $LaNiO_2$ has a comparatively small $\Delta I_T$ across the measured energy range, whereas $\Delta I_T$ increases with Ca doping. In the pre-edge region (Region I, highlighted in yellow), which contains features "a" (Ni3$d$–O2$p$ hybridization) and "b" (the superconductivity-associated band), $\Delta I_T$ increases with doping and reaches a positive saturation near $x = 0.23$, indicating stronger absorption at 300 K than at 30 K. For higher doping levels ($x > 0.23$), $\Delta I_T$ decreases, but remains noticeably positive. In Region II (highlighted in green), where feature "c" associated with La5$d$/4$f$–O2$p$ hybridization appears, $\Delta I_T$ becomes substantially larger for x⩾0.10 and stays elevated, although without a clear monotonic doping dependence. At higher photon energies, beyond ~535.1 eV (Region III, highlighted in blue), the temperature differential, $\Delta I_T$, is comparatively small and does not show a clear trend. These observations indicate that O2$p$ hybridization with both Ni- and La-derived states is generally stronger at room temperature than at low temperature, with the most pronounced temperature dependence occurring in the low-energy pre-edge region.

### Ni L-edge XAS spectra

In addition to the O K-edge, we performed XAS measurements at the Ni $L_3$-edge. The main absorption peak at ~854.3 eV corresponds to the $2p^6{}_{3/2}3d^9 \rightarrow 2p^5{}_{3/2}3d^{10}$ transition[28,31,32]. Room-temperature $L_3$-edge spectra for different Ca concentrations are shown in Fig. 3(a) (see Fig. S3(a) for the corresponding low-temperature spectra).

At both 300 K and low temperature, the $L_3$-edge intensity increases with Ca doping and reaches a maximum as the system approaches the superconducting regime, peaking near $x = 0.18$ ($La_{0.82}Ca_{0.18}NiO_2$). With further doping, the $L_3$-edge intensity decreases, but remains higher than in undoped $LaNiO_2$. This behaviour is summarized by the peak-intensity ratio at 300 K, $I_{NiL}(x)/I_{NiL}(x = 0.18)$, plotted in Fig. 3(b). When the Ni $L_3$ intensity trend is overlaid with the doping-dependent phase diagram of infinite-layer $La_{1-x}Ca_xNiO_2$ thin films, a parallel evolution is evident from the undoped state to the onset of superconductivity and into the near-optimal regime ($0 < x < 0.18$). This is consistent with Ca substitution progressively hole-doping the system and increasing the effective Ni valence from $Ni^{1+}$ toward $Ni^{(1+x)+}$. However, the enhancement of superconducting behaviour is confined to a finite doping range. Beyond the optimal regime ($x > 0.23$), $T_c$ decreases and superconductivity disappears for $x > 0.3$.

This evolution differs from the doping dependence seen at the O K-edge, particularly when considering the emergence and strengthening of feature "b" in the pre-edge region (Region I in Fig. 1(b)). The integrated-area ratio of Region I at 300 K, $I_{OK}(x)/I_{OK}(x=0.23)$, overlaid in Fig. 3(b), peaks at $x = 0.23$. Similar trends are obtained at 30 K (Supplementary Fig. S4), where corresponding overlays of the Ni L-edge intensity and O K-edge integrated-area ratios show consistent doping-dependent behaviour. The O K-edge and Ni L-edge results indicate a systematic redistribution of low-energy spectral weight with Ca doping, from more Ni3$d$–dominated character toward increasingly O2$p$–hybridized states. Notably, the reduction of $T_c$ beyond the near-optimal regime occurs in the doping range where the orbital character changes most rapidly. Furthermore, the doping-dependent Hall coefficient reported in Ref. 36 provides additional evidence that the electronic evolution cannot be captured by carrier-density changes alone. In particular, the sign change of $R_H$ in the overdoped region ($x > 0.23$) coincides with the weakening of Ni3$d$-derived spectral weight and the growing dominance of O-hybridized states. This concurrence supports an orbital-selective interpretation of both the transport anomalies and the superconducting phase diagram, going beyond a purely phenomenological multiband description.

**Ni $L_3$-edge Linear Dichroism**

We further examined the polarization dependence of the Ni $L_3$-edge via XLD at both high and low temperatures. In addition to normal-incidence measurements, spectra were collected at a grazing incidence angle (20°). At grazing incidence, the $L_3$-edge intensity again peaks near $x = 0.18$ at both temperatures; representative room-temperature spectra are shown in Fig. 3(c) (the corresponding grazing-incidence intensity ratios are shown in Fig. S5).

Interestingly, unlike Nd-based infinite-layer nickelates ($Nd_{1-x}Sr_xNiO_2$), which exhibit pronounced dichroism between in-plane and out-of-plane polarizations[31,35], the La-based films show only weak dichroism. The in-plane component is only slightly larger than the out-of-plane component across the measured doping and temperature ranges (room-temperature XLD shown in Fig. 3(d)). This suggests that the unoccupied Ni 3$d$-states have comparable in-plane and out-of-plane character, implying a relatively small energy splitting between the in-plane $3d_{x2-y2}$ and out-of-plane $3d_{z2}$ components in the square-planar $NiO_2$ environment.

**Conclusions**

We have shown that O2$p$–Ni3$d$ hybridization remains robust across the doping series of infinite-layer $La_{1-x}Ca_xNiO_2$. In the superconducting regime, an additional spectral feature appears in the O K-edge pre-edge region, indicating a pronounced modification of the low-energy unoccupied states. Across the superconducting window ($0.18 \leq x \leq 0.27$), we observe a redistribution of low-energy spectral weight: Ni3$d$-dominated contributions decrease while O2$p$-hybridized states increase, most rapidly near $x \approx 0.20$-0.23. This crossover coincides with a sign change in the Hall coefficient, $R_H$, pointing to an orbital-selective reorganization of the low-energy electronic structure that goes beyond simple carrier modulation. At higher doping ($x > 0.23$), the continued growth of O-hybridized character tracks the suppression of $T_c$, linking the downturn of the superconducting dome to a hybridization-driven electronic evolution.

Temperature- and doping-dependent Ni $L_3$-edge measurements further reveal behaviour that differs from the Nd-based analogue $Nd_{1-x}Sr_xNiO_2$, particularly in the weak XLD response, which implies a smaller anisotropy between in-plane and out-of-plane unoccupied $3d$ states. Overall, these findings place Ni–O covalency and its doping evolution at the centre of the electronic-structure changes across the superconducting dome in La-based infinite-layer nickelates, refine the phase diagram by connecting transport anomalies to orbital redistribution, and suggest hybridization control as a practical route for tuning superconducting behaviour.

**Acknowledgments**

This work was supported by the Ministry of Education (MOE), Singapore, under its Tier-2 Academic Research Fund (AcRF), Grants No. MOE-T2EP50123-0013 and MOE-T2EP50124-0003, and by the National Research Foundation (NRF) Investigatorship Award on ”Quantum design of superconductivity and correlated phases in layered complex oxides”. S.W.Z., Z.S.L., and H.J.L. acknowledge the RIE2025 MTC Individual Research Grants (M22K2c0084), NRF Competitive Research Program (NRF-CRP28-2022-0002), Career Development Fund (C210812020) and Central Research Fund from the Agency for Science, Technology and Research (A*STAR) for the funding support. X.M.Y. acknowledges financial support by National Natural Science Foundation of China (Grant No.12374378). The authors would like to acknowledge the Singapore Synchrotron Light Source for providing the facility necessary for conducting the research. The Laboratory is a National Research Infrastructure under the National Research Foundation, Singapore. Any opinions, findings, and conclusions or recommendations expressed in this material are those of the author(s) and do not reflect the views of National Research Foundation, Singapore.

**Author contributions**

C.S.T., S.W.Z., and A.A. conceived the project. S.W.Z., X.G., Z.Y.L., Z.S.L. and S.P. prepared the thin films and conducted the electrical measurements. C.S.T., C.Z.D., and M.B.H.B. conducted the XAS measurements. X.M.Y. and C.J.L. provide insight to manuscript. C.S.T., S.W.Z., H.J.L., and A.A. wrote the manuscript with contributions from all authors. All authors have discussed the results and the interpretations.

**Competing interests**

The authors declare no competing interests.

## Figures and Captions

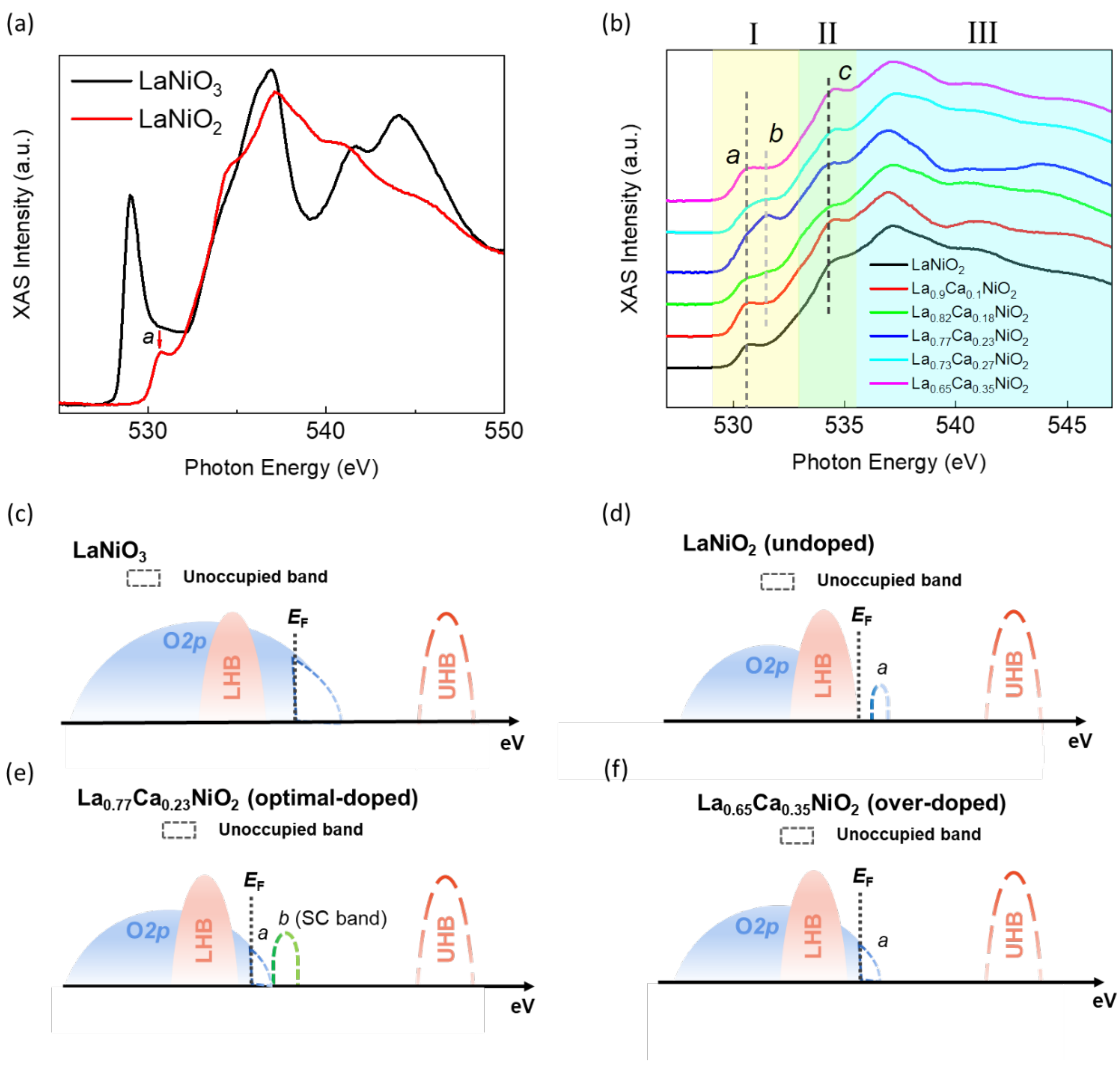

**Fig. 1: Doping-dependent O K-edge spectra**. (a) Representative O K-edge spectra comparing perovskite phase $LaNiO_3$ and reduced infinite-layer phase $LaNiO_2$. Feature "a" detected in the infinite-layer phase denoted by red arrow at ~530.7 eV. (b) Doping-dependent infinite-layer phase $La_{1-x}Ca_xNiO_2$ from the underdoped to the over-doped phase. Features "a", "b" and "c" identified with dashed lines as visual guides. Schematic band structure of (c) perovskite phase $LaNiO_3$, (d) undoped $LaNiO_2$, (e) optimal-doped $La_{0.77}Ca_{0.23}NiO_2$, and (f) over-doped $La_{0.65}Ca_{0.35}NiO_2$.

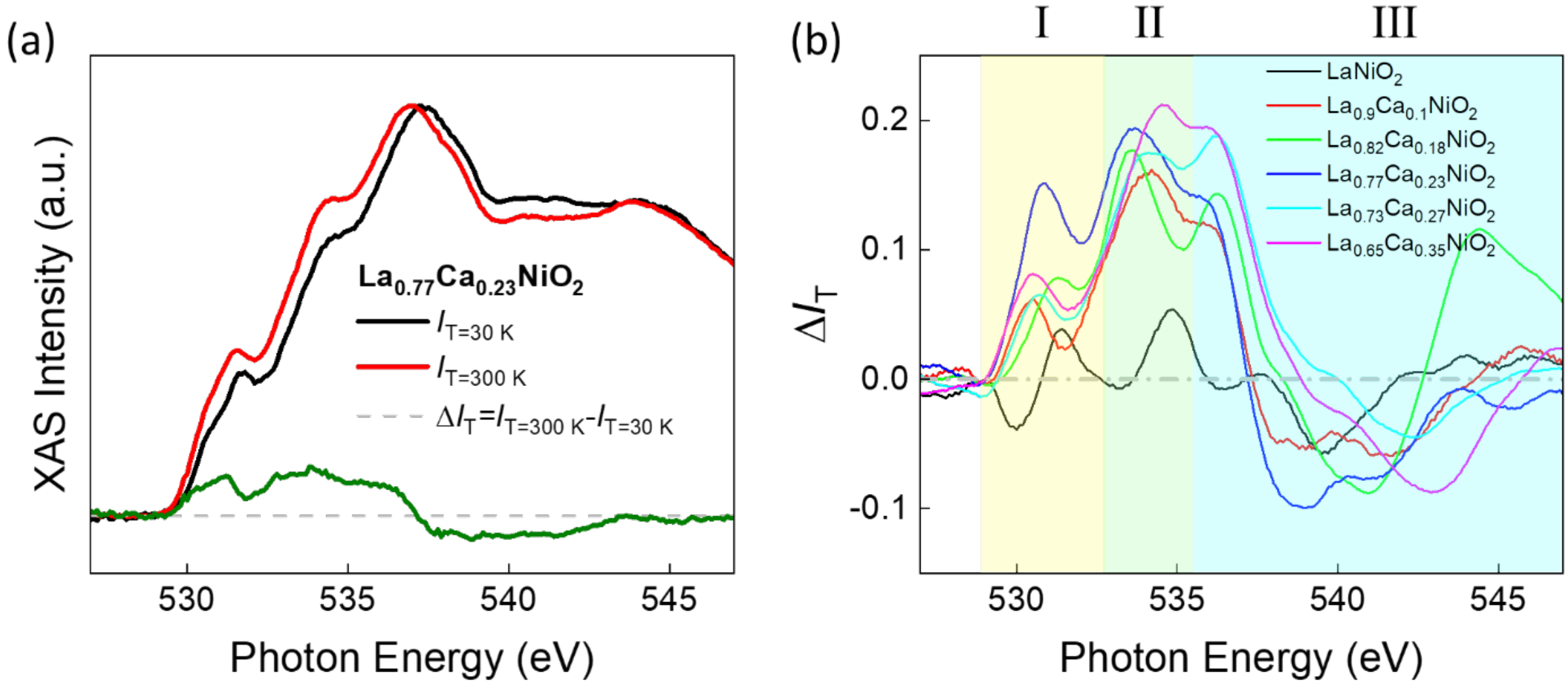

**Fig. 2: Temperature-dependent O K-edge spectra.** (a) Representative XAS spectra of $La_{0.77}Ca_{0.23}NiO_2$ at 30 and 300 K overlaid with the corresponding temperature differential, $\Delta I_T$ [where $\Delta I_T = I_{T=300\text{ K}} - I_{T=30\text{ K}}$]. (b) comparing the O K-edge temperature differential, $\Delta I_T$, across the doping concentration. The colour shading represents the photon energy regions where features "a", "b" (in the pre-peak region), "c" and beyond are observed.

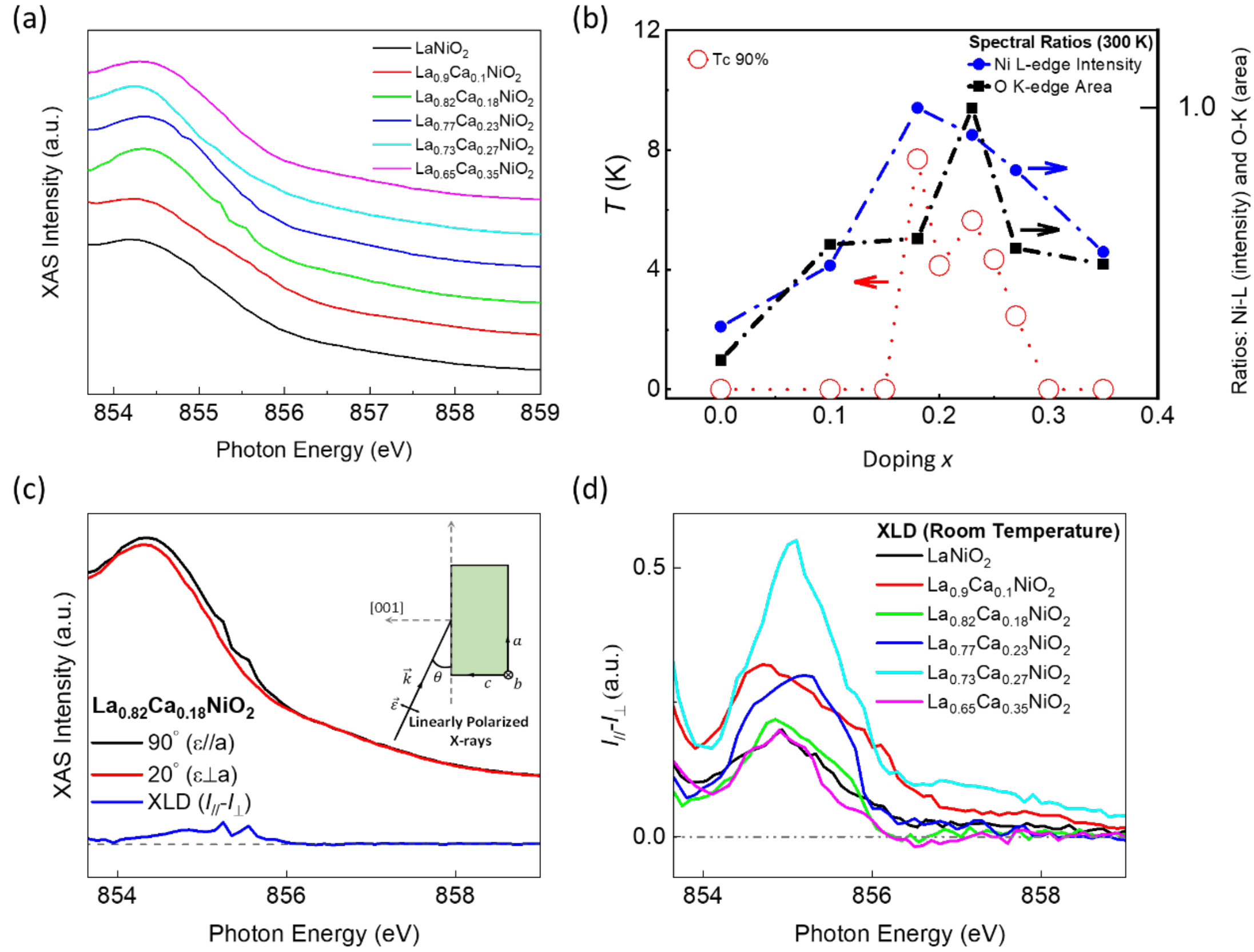

**Fig. 3: Ni L3-edge XAS spectra.** (a) Ni $L_3$-edge from the undoped through the over-doped phase at room temperature. (b) Ni $L_3$-edge peak intensity ratio (in blue), $I_{\mathrm{NiL}}(x)/I_{\mathrm{NiL}}(x=0.18)$, and O K-edge integrated area ratio of Region I (in black), $I_{\mathrm{OK}}(x)/I_{\mathrm{OK}}(x=0.23)$, respectively at 300 K. The intensity ratio is overlaid with the critical temperature as a function of doping level $x$ for $La_{1-x}Ca_xNiO_2$. The $T_{c,90\,\%\,\mathrm{R}}$ data (in red circles) is defined as the temperature at which the resistivity drops to 90% of the value at 10 K (the onset of the superconductivity). The $T_c$ data of $La_{1-x}Ca_xNiO_2$ are adapted from Ref. 36. The Ni $L_3$-edge peak intensity ratio and the O K-edge integrated area ratio are not in the same scale. (c) Representative Ni $L_3$-edge spectra of $La_{0.82}Ca_{0.18}NiO_2$ alongside the intensity difference between the in-plane and out-of-plane components [where XLD, $\Delta I = I(\varepsilon \| a) - I(\varepsilon \perp a)$]. Inset: Direction of photon polarization $\varepsilon$ and incident angle. (d) Room temperature XLD of $La_{1-x}Ca_xNiO_2$ across the doping range.

# Supplementary Information

## Ruling out the signals from the STO substrate layer

Near edge XAS (NEXAFS) measurements, particularly when employing the total electron yield method, are characterized by a penetration depth of roughly 5 nm.[32,33] This relatively shallow penetration allows for detailed surface-sensitive analyses of materials, making it an invaluable tool for studying thin films and surface phenomena. The technique is widely used in various fields such as material science, chemistry, and physics due to its ability to provide insights into the electronic structure and chemical state of the surface atoms of a sample. Of particular interest in these measurements is the observation of pre-peaks (Fig. 1), which are often indicative of specific electronic transitions or structural features within the material. Notably, even when the O K-edge XAS experiment is conducted at a grazing incident angle of 20$^{o}$ (Fig. S2), these pre-peaks remain visible at both the low- and high-temperature regimes. This persistence suggests that the pre-peak phenomena are not merely surface artefacts or from the STO substrate as reported by previous studies.[16,20] Instead, these are intrinsic to the infinite-layer electronic structure and composition. The visibility of the pre-peak under these conditions reinforces the notion that it has a significant attribution to certain electronic states or structural characteristics within the sample, warranting further investigation.

## Supplementary Figures

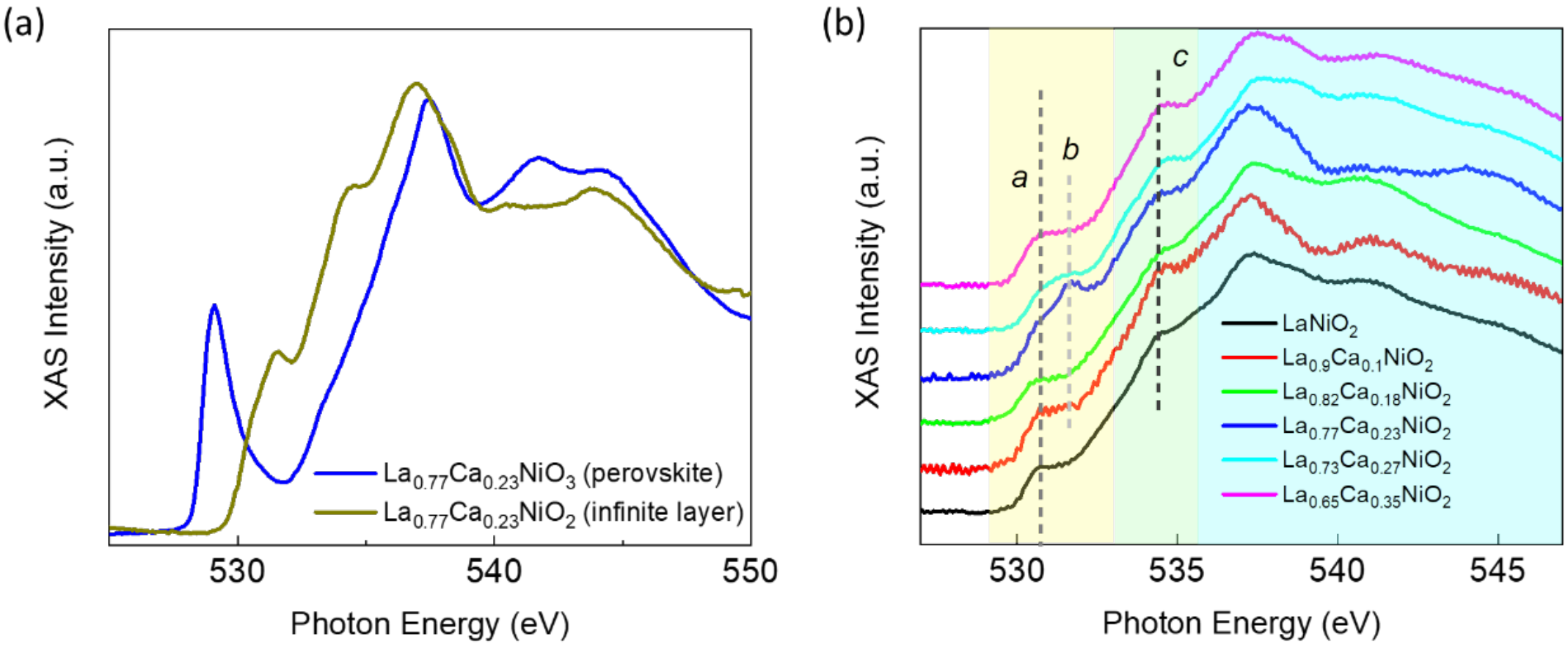

**Fig. S1.** (a) Representative O K-edge XAS spectra of the perovskite phase $La_{0.77}Ca_{0.23}NiO_3$ and infinite-layer phase $La_{0.77}Ca_{0.23}NiO_2$ at room temperature. (b) Overlay of O K-edge XAS spectra of infinite-layer phase $La_{1-x}Ca_xNiO_2$ across different doping concentration at low temperature (30 K).

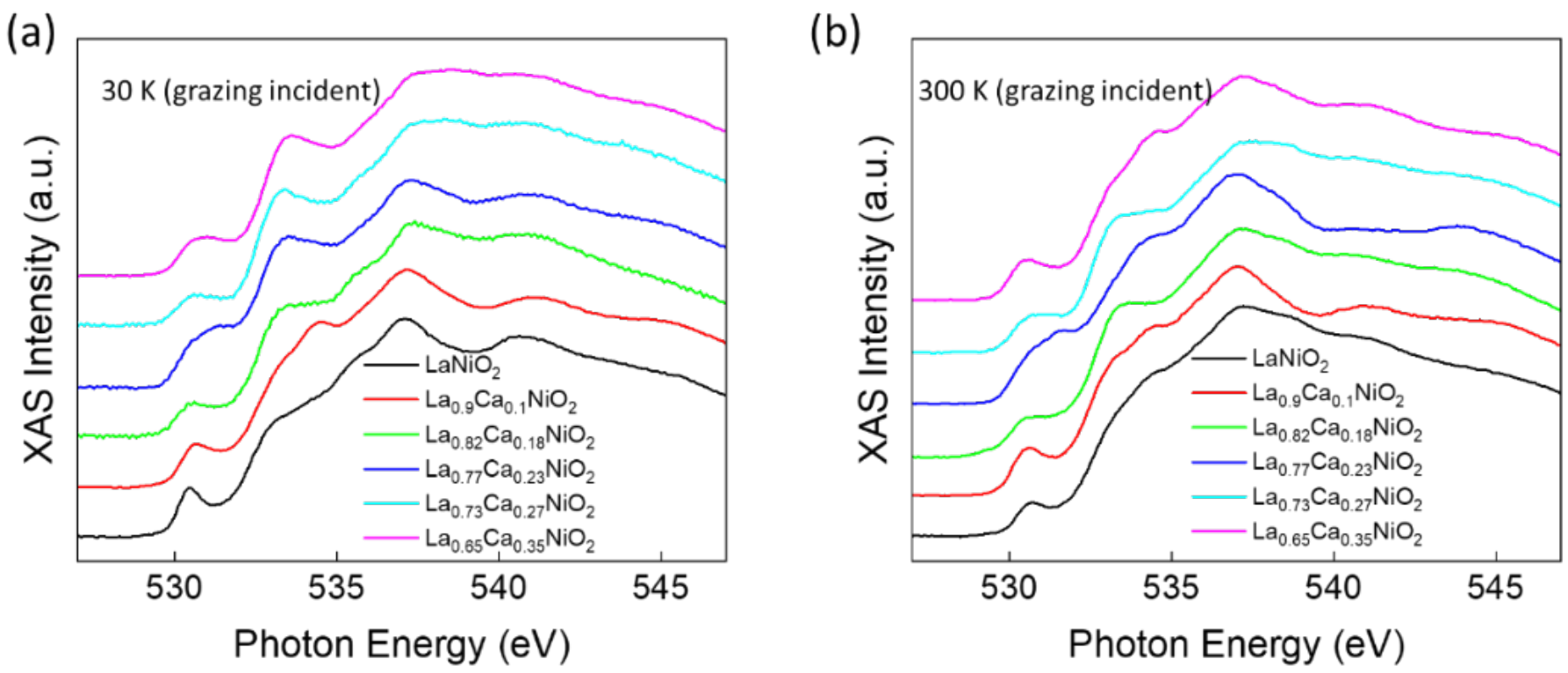

**Fig. S2.** O K-edge XAS spectra conducted at incident grazing angle (20°) of infinite-layer phase $La_{1-x}Ca_xNiO_2$ across different doping concentration at (a) 30 K, and (b) 300 K.

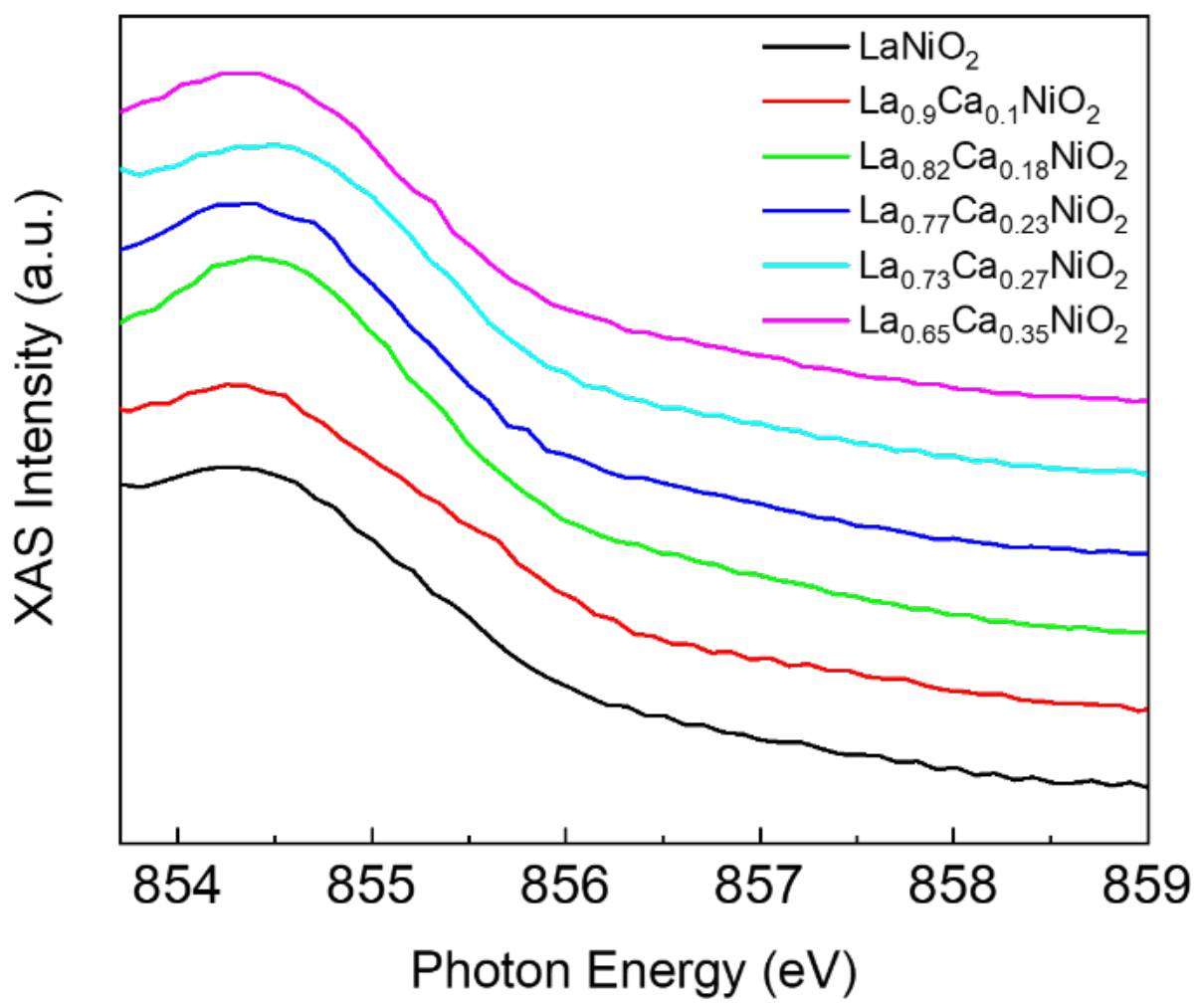

**Fig. S3.** Overlay of the normal incident (90°) Ni $L_3$-edge spectra across different doping concentration at low temperature (30 K).

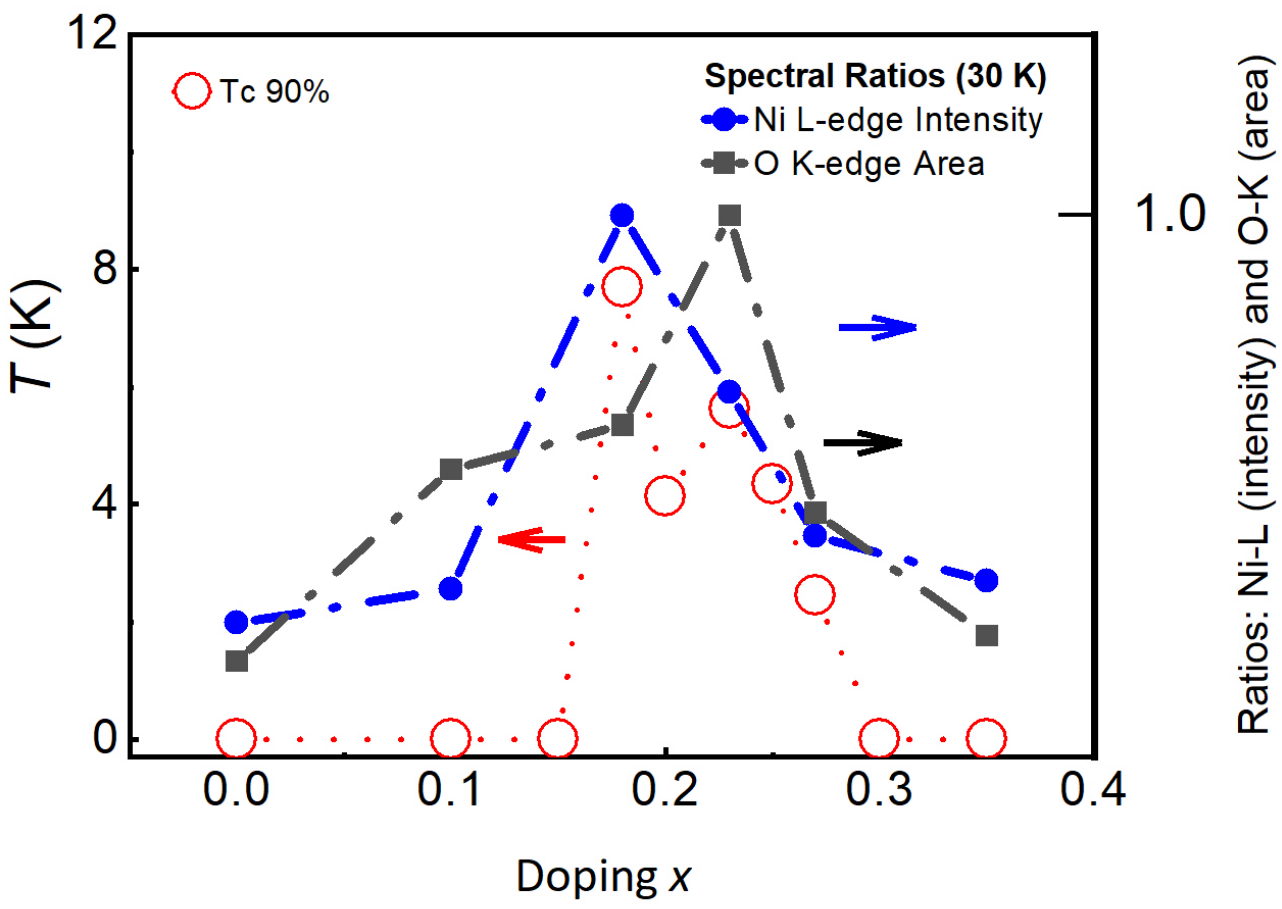

**Fig. S4.** Ni $L_3$-edge peak intensity ratio (in blue), $I_{NiL}(x)/I_{NiL}(x = 0.18)$, and O K-edge integrated area ratio of Region I (in black), $I_{OK}(x)/I_{OK}(x{=}0.23)$, respectively at 30 K with reference to the vertical axis on the right. For comparison, these intensity ratios are overlaid with the $T_{c,90\%}$ data (in red circles) as a function of doping level $x$ for $La_{1-x}Ca_xNiO_2$, adapted from Ref. 36. The $T_{c,90\%}$ data (in red circles) is defined as the temperature at which the resistivity drops to 90% of the value at 10 K (the onset of the superconductivity). The Ni $L_3$-edge peak intensity ratio and the O K-edge integrated area ratio are not in the same scale.

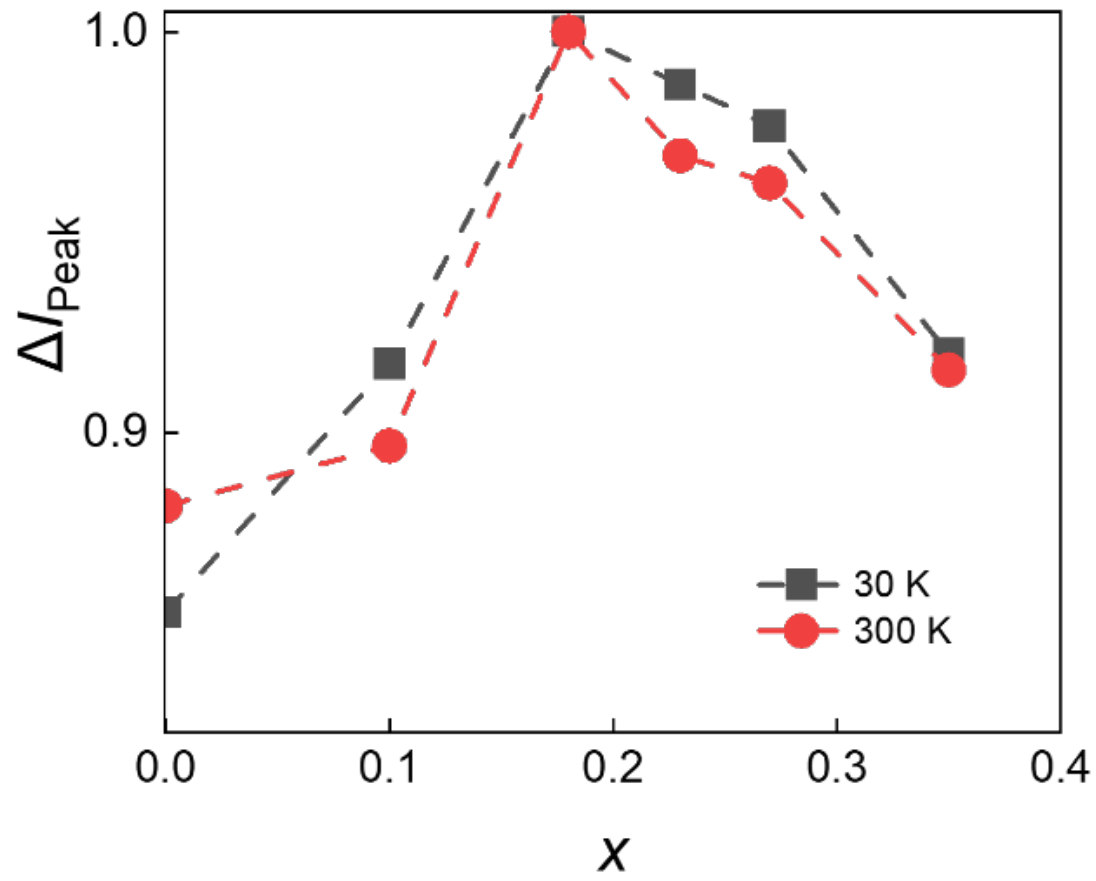

**Fig. S5.** Grazing incident angle (20°) Ni $L_3$-edge peak intensity ratio, defined by $I(x)/I(x = 0.18)$, at 30 and 300 K, respectively, which register similar intensity variations as the normal incidence data (90°) in main text Fig. 3(b).